\documentstyle[12pt,epsfig]{article}
\newcommand{\newc}{\newcommand}
\newcommand{\be}[1]{\begin{equation} \label{(#1)}}
\newcommand{\ee}{\end{equation}}
\newcommand{\ba}[1]{\begin{eqnarray} \label{(#1)}}
\newcommand{\ea}{\end{eqnarray}}

\def\pmb#1{\setbox0=\hbox{#1}%
  \kern-.015em\copy0\kern-\wd0
  \kern.03em\copy0\kern-\wd0
  \kern-.015em\raise.0233em\box0 }
\def \sw {\sin\!\theta^{}_W }

\def \sw2 {\sin^2\!\theta^{}_W }

\def \s2bt {\sin\!2\beta }

\newc{\Rslash}{\not R}
\newc{\superhu}{\hat H_u}
\newc{\superhd}{\hat H_d}
\newc{\superl}{\hat L}		
\newc{\superr}{\hat R}
\newc{\superec}{\hat {e}^c}
\newc{\superq}{\hat Q}
\newc{\superu}{\hat U}
\newc{\superd}{\hat D}
\newc{\superuc}{\hat {u}^c}
\newc{\superdc}{\hat {d}^c}
\newc{\Lsoft}{{\cal L}_{\rm soft}}
\newc{\hc}{{\rm H.c.}}
\newc{\tildeQ}{\widetilde Q}
\newc{\tildeU}{\widetilde U}
\newc{\tildeD}{\widetilde D}
\newc{\tildeL}{\widetilde L}
\newc{\tildeuc}{\widetilde {u}^c}
\newc{\tildedc}{\widetilde {d}^c}
\newc{\tildeec}{\widetilde {e}^c}
\newc{\tildenu}{\widetilde \nu}
\newc{\onehalf}{\frac{1}{2}}
\newc{\gluino}{{\tilde g}}
\newc{\mgluino}{m_{\gluino}}
\newc{\mzero}{m_0}
\newc{\mz}{m_Z}
\newc{\mw}{m_W}
\newc{\alphas}{\alpha_s}
\newc{\tanb}{\tan\beta}
\newc{\hinot}{\widetilde H^0_2}
\newc{\hinob}{\widetilde H^0_1}
\newc{\bino}{\widetilde B^0}
\newc{\wino}{{\widetilde W^0_3}}
\newc{\msl}{m_{\widetilde l}}
\newc{\msel}{m_{\tilde e}}
\newc{\msq}{m_{\tilde q}}
\newc{\msf}{m_{\tilde f}}
\newc{\mev}{{\rm\,MeV}}
\newc{\gev}{{\rm\,GeV}}
\newc{\tev}{{\rm\,TeV}}

\begin{document}

\begin{center}
{\Large \bf Test of Physics beyond the Standard Model in Nuclei}\\

\bigskip {Amand Faessler\footnote{
Supported by the European Union under 
the network contract CT93-0323 and the Deutsche Forschungsgemeinschaft 
Fa67/17-1 and Fa67/19-1 } 
and Fedor \v Simkovic\footnote{{\it On leave from:}
 Department of Nuclear physics, Comenius University, 
SK--842 15 Bratislava, Slovakia
}}\\
\bigskip 

{\it Institute f\"ur Theoretische Physik  der Universit\"at
T\"ubingen,\\[0pt]
Auf der Morgenstelle 14, D-72076 T\"ubingen, Germany}\\[0pt]
\end{center}
\bigskip

\begin{abstract}
The modern theories of Grand Unification (GUT)
and supersymmetric (SUSY) extensions of standard model (SM) 
suppose that the conservation laws of the SM may be
violated to some small degree. 
The nuclei are well-suited as a laboratory to test
fundamental symmetries and fundamental interactions 
like lepton flavor (LF) and lepton number (LN) conservation.  
A prominent role between experiments 
looking for LF and total LN violation play 
yet not observed processes of neutrinoless double beta decay
($0\nu\beta\beta$-decay). The GUT's and SUSY models offer a variety
of mechanisms which allow $0\nu\beta\beta$-decay to occur. They are based
on mixing of Majorana neutrinos  and/or  
R-parity violation hypothesis. 
Although the $0\nu\beta\beta$-decay has not been seen it is
possible to extract from the lower limits of the lifetime upper limits 
for the effective electron Majorana neutrino mass, effective right handed weak
interaction parameters, the effective Majoron coupling constant,
R-parity violating SUSY parameters etc.  

A condition for obtaining reliable limits for these fundamental
quantities is that the nuclear matrix elements governing this process
can be calculated correctly. The nuclear structure wave functions can
be tested by calculating the two neutrino double beta decay 
($2\nu\beta\beta$-decay) for which we have experimental data and not only
lower limits as for the $0\nu\beta\beta$-decay.
For open shell nuclei the method of choice has been the 
Quasiparticle Random Phase Approximation, which treats Fermion 
pairs as bosons. It has been found that by extending the QRPA including
Fermion commutation relations better agreement with $2\nu\beta\beta$-decay
experiments is achieved. This increases also the reliability of 
conclusions from the upper limits on the  $0\nu\beta\beta$-decay
transition probability.

In this work the limits on the LN violating parameters
extracted from current $0\nu\beta\beta$-decay experiments are
listed. Studies in respect to future $0\nu\beta\beta$-decay experimental 
projects are also presented. 
\end{abstract}

\section{Introduction}
\label{sec:level1}

The standard model (SM) represents the simplest and most economical theory
which describes jointly weak and electromagnetic interactions. 
It describes well all terrestrial experimental results known today.
Nevertheless the SM cannot be considered as the ultimative 
theory of nature and is likely to describe the effective interaction
at low energy of an underlying more fundamental theory. The 
supersymmetry (SUSY), 
which is one of the fundamental new symmetries of nature,
 is believed to be next step beyond the successful SM. 
The supersymmetry is the symmetry between fermions and bosons, which
has to be broken in order to explain the phenomenology of the
elementary particles and their superpartners.
It is the only known symmetry which can stabilize the 
elementary Higgs boson mass with respect to otherwise uncontrollable 
radiative corrections. 
The minimal supersymmetric model (MSSM), that leads to the SM 
at low energies has been the subject of extensive investigations.

Both the SM and the MSSM leads to zero mass for the neutrinos.
In view of the observed results on solar 
(Homestake \cite{cle95}, Kamiokande \cite{hir91}, 
Gallex \cite{gal96} and SAGE \cite{abd94})
and atmospheric neutrinos 
(IMB \cite{bec95}, Soudan 2 \cite{goo96}, MACRO \cite{macro} and
Super-Kamiokande \cite{kam98}) it is more
appropriate to consider the extensions of the SM and  the MSSM that can 
lead to neutrino masses.
Neutrino masses either require the existence of right-handed neutrinos
or require violation of the lepton number (LN)  so that Majorana masses are
possible. So, one is forced to go beyond the minimal models again, 
whereby LF and/or LN violation can be allowed 
in the theory. 
A good candidate for such a theory is  the left-right symmetric 
model of Grand Unification (GUT) inaugurated by
Salam, Pati, Mohapatra and Senjanovi\'c \cite{rlt} (especially models based
on SO(10) which have first been proposed by
Fritzsch and Minkowski \cite{fri}) and its supersymmetric version
\cite{srlt}. The left-right symmetric models,  
representing generalization of the 
$SU(2)_L \otimes U(1)$ SM, predict not only  
that the neutrino is a Majorana particle, that means it is up to a 
phase identical with its antiparticle, but automatically 
predict the neutrino has a mass and a weak right-handed interaction.
The basic idea behind grand unified models is an extension of the
local gauge invariance from quantum chromodynamics (SU3) involving
only the colored quarks also to electrons and neutrinos. We note that
the non-supersymmetric left-right models suffer from the hierarchy
problem. 

The expectations arising from GUT's and theirs SUSY versions
are that the conservation laws of the 
SM, e.g. LN conservation,  may be violated to some small degree. 
In the left-right symmetric models the LN conservation is
broken by the presence of the Majorana neutrino mass.
The LN violation is also inbuilt in those SUSY theories 
where R-parity, defined as 
$R_p = (-1)^{3B+L+2S}$ (S, B, and L are the spin, baryon and lepton
number, respectively) is not a conserved quantity anymore.
The conservation of LN is among the most 
stringently tested laws of physics nowadays.
The nuclei are well-suited as laboratory to test this fundamental
symmetry due to the fact that a variety of quantum numbers is available 
as initial and final states. The neutrinoless double beta decay
($0\nu\beta\beta$-decay),
\begin{equation}
(A,Z) \rightarrow (A,Z+2) + 2e^-,
\label{eq:1}
\end{equation}
which involves the emission of two electrons and no neutrinos,
has been long recognized as a powerfool tool to study 
the LN conservation. The $0\nu\beta\beta$-decay takes place 
only if the neutrino is a Majorana particle 
with non-zero mass \cite{schv}. The GUT's and R-parity violating
SUSY models offer a plethora of the $0\nu\beta\beta$-decay 
mechanisms triggered 
by exchange of neutrinos, neutralinos, gluinos, leptoquarks etc.
\cite{fae98,suh98}. If one assumes that one mechanism at a time
dominates, the half-life of the $0\nu\beta\beta$-decay can 
be written as
\begin{equation}
( T^{0\nu}_{1/2})^{-1} = |LNV|^2 \sum_i P_i^{0\nu} G_i^{0\nu},
\label{eq:2}
\end{equation}
where LNV is some effective LN violating parameter,
$P_i^{0\nu}$ is the real part of the product of two
nuclear matrix elements governing
the $0\nu\beta\beta$-decay and $G_i^{0\nu}$ is the integrated kinematical 
factor. The sum over $i$ runs over different phase space integrals
weighted by a corresponding product of nuclear matrix elements.
There are different LNV parameters, e.g.
the effective electron-neutrino mass, effective right handed weak
interaction parameters, effective Majoron coupling constant and
R-parity violating SUSY parameters, which incorporate elements
of the fundamental interaction of Majorana neutrinos and/or 
R-parity violating interaction of SUSY particles (see e.g. the recent
review articles \cite{fae98,suh98}).  The value of these parameters
can be determined in two ways: i) One can extract upper bound on the
LNV parameter from the best presently available experimental lower 
limit on the half life of the $0\nu\beta\beta$-decay 
$T^{0\nu -exp}_{1/2}$
after calculating the corresponding nuclear matrix elements. 
ii) One can use the phenomenological constraints imposed by other
experiments, e.g. those looking for the neutrino oscillation, to
evaluate the  LNV parameter explicitly, which further can be compared 
with the extracted one. 
The $0\nu\beta\beta$-decay constraints on LNV
parameters must be taken into account by the theoreticians, when
they build new theories of grand unification.

At present the searches for  $0\nu\beta\beta$-decay are pursued 
actively for different nuclear isotopes, e.g. $^{76}Ge$
(Heidelberg-Moscow coll. \cite{bau99} and IGEX coll. \cite{aal99}),
$^{100}Mo$ (NEMO \cite{xsa98},  and ELEGANTS \cite{eji96}),
$^{116}Cd$ (the INR exper. \cite{dane95} and the NEMO exper. \cite{arn96}), 
$^{130}Te$ \cite{alle98} and $^{136}Xe$ 
(the Gotthard Xe exper. \cite{bus98}). The sensitivity
of a given isotope to the different LN violating signals
is determined by the value  of the corresponding nuclear matrix element 
connecting the ground state of the initial and final nuclei with
$J = 0^+$ and the  value of the kinematical factor determined by the
energy release for this process. In order to correctly interpret the 
results of $0\nu\beta\beta$-decay experiment, i.e. to obtain
qualitative answers for the LN violating parameters,
the mechanism of nuclear transitions has to be understood. 
The $0\nu\beta\beta$-decay nuclear systems of interest are medium and 
heavy open shell nuclei with complicated nuclear structure.
To test our ability to evaluate the nuclear matrix elements that 
govern the decay rate, it is desirable to describe the 
two-neutrino double beta decay ($2\nu\beta\beta$-decay)
allowed in the SM:
\begin{equation}
(A,Z) \rightarrow (A,Z+2) + 2e^- + 2\tilde{\nu}.
\label{eq:3}
\end{equation}
We note that each mode of the double beta decay requires 
the construction of the same many-body nuclear structure wave functions.

A variety of nuclear techniques have been used in attempts to calculate
$2\nu\beta\beta$-decay matrix elements, which have been reviewed 
recently in Refs. \cite{fae98,suh98}. Especially the  
Quasiparticle Random Phase Approximation (QRPA) and its extensions
have been found  to be powerful models, 
considering their simplicity, to describe nuclear matrix elements, 
which require the summation over a complete set of intermediate
nuclear states. The recent $2\nu\beta\beta$-decay calculations 
\cite{toi95,simn96,ves97} including the schematic ones \cite{sam97,fed99}
manifest that the inclusion of the Pauli exclusion principle (PEP) in the QRPA 
improves the predictive power of the theory giving  more reliable
prediction of the $2\nu\beta\beta$ decay probability.

In this contribution we present the upper limits on some effective 
LN violating parameters extracted  from the current 
experimental limits of the $0\nu\beta\beta$-decay lifetime 
for $A=76$, 82, 96, 100, 116, 128, 130, 136 and 150 isotopes, 
which are quantities of fundamental importance.  A discussion
in respect to the sensitivity of a given $0\nu\beta\beta$-decaying isotope
to the different LN violating signals is presented.
Some related nuclear physical aspects as well as studies in
respect to future $0\nu\beta\beta$-decay experiments are addressed.

\section{Neutrinoless double beta decay}
\label{sec:level2}

\subsection{Majorana neutrino mixing mechanisms}

The presently favored models of grand unification
are left-right symmetric models \cite{rlt}. They
contain left- and hypothetical right-handed vector bosons 
$W^{\pm}_L$ and $W^{\pm}_R$. 
The vector bosons mediating the left and right-handed interaction are mixed 
if the mass eigenstates $W^{\pm}_1$ and $W^{\pm}_2$ 
are not identical with the weak eigenstates, which
have a definite handedness.:
\begin{eqnarray}
W^{\pm}_1 & = &  ~~~\cos \zeta \cdot W^{\pm}_L + \sin \zeta \cdot
W^{\pm}_R \cr
W^{\pm}_2 & = & - \sin \zeta \cdot W^{\pm}_L + \cos\zeta \cdot
W^{\pm}_R.
\label{eq:4}
\end{eqnarray}
$\zeta$ is the mixing angle of the vector bosons.
The left-right symmetry is broken
since the vector bosons $W^{\pm}_1$ and $W^{\pm}_2$ obtain
different masses by the Higgs mechanism. 
Since we have not seen a
right-handed weak interaction the mass of the heavy, mainly
``right-handed'' vector boson must be much larger than the mass of the
light
(81 GeV) vector boson, which is responsible for the left-handed force.

\begin{figure}[t]
\centerline{\epsfig{file=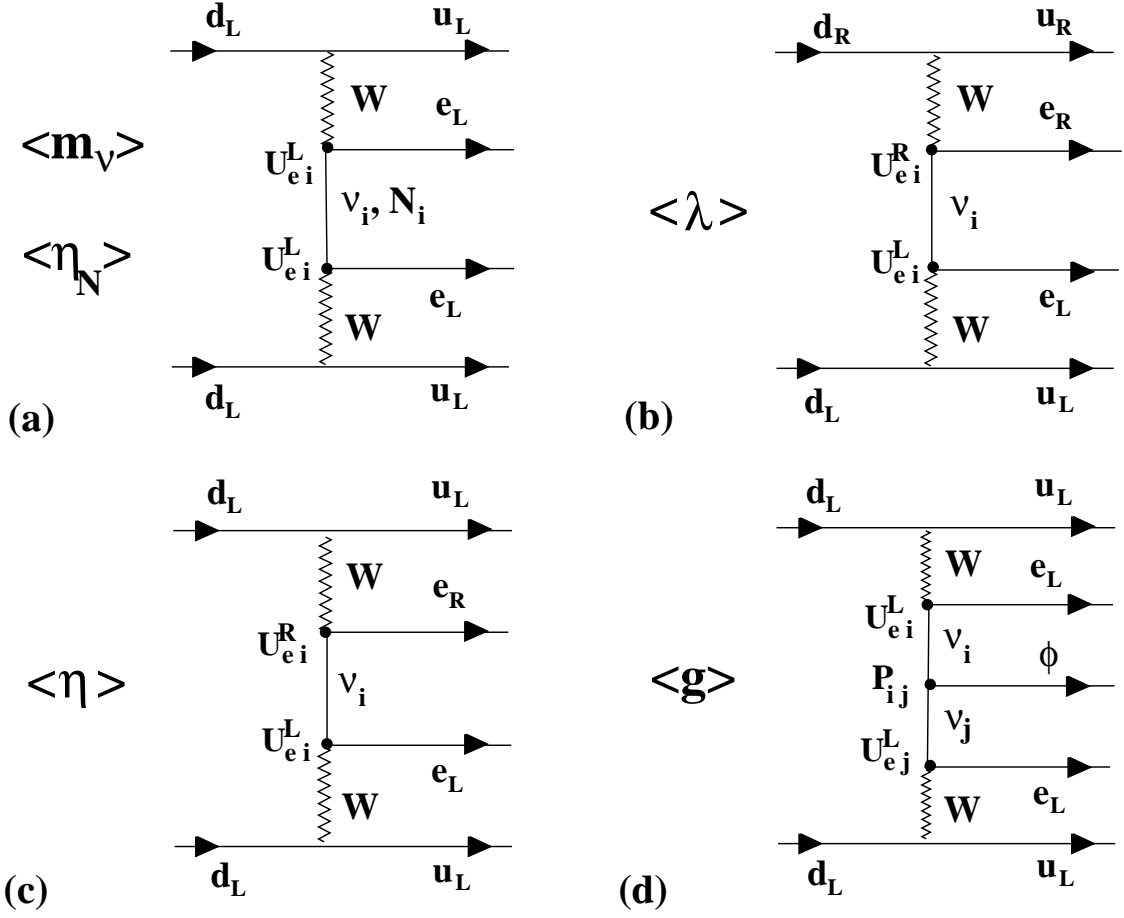,height=12.0cm}}
\caption{Mechanisms for $0\nu\beta\beta$-decay associated with 
the exchange of a Majorana neutrino: (a) the light and the heavy 
neutrino mass mechanism, (b), (c) right handed current mechanisms,
(d) the Majoron mechanism. 
The following notation is used: $u_{L(R)}$, $d_{L(R)}$ and $e_{L(R)}$
are left- (right-) handed u-quark, d-quark and electron, respectively.
$W$ is vector boson (light or hypothetical heavy) and $\nu_i$ (i=1,2...)
is the Majorana neutrino. 
}
\label{fig.1}
\end{figure}

The weak interaction Hamiltonian must now be generalized.
\begin{eqnarray}
H_W & \approx & {\frac{G_F}{\sqrt{2}}} \left[ 
({J_L} \cdot {j_L}) 
+ tg \zeta ({J_R} \cdot {j_L})
+ tg \zeta ({J_L} \cdot {j_R}) +
{ \left( {\frac{M^W_1}{M^W_2}} \right) }^2 ({J_R} \cdot {j_R}) 
\right] + h.c., 
\nonumber \\
{J_L}/{J_R} & = & \bar{\psi}_p \gamma_\mu
(g_V \mp g_A \gamma_5) \psi_n, ~~~~~
{j_L}/{j_R} = \bar{e} \gamma_{\mu} (1 \mp \gamma_5) 
\nu_{e L/R }, 
\label{eq:5}
\end{eqnarray}
where  $g_V = 1$ and $g_A = 1.25$. $M^W_1 $ and
$M^W_2$ are the light and heavy vector boson masses, respectively. 
The capital ${J_L}$ and ${J_R}$ 
indicate the hadronic left- and right-handed
currents changing a neutron into a proton, respectively. 
The lower case ${j_l}$ (${j_r}$) is the
left (right) handed leptonic current  which 
create an electron (or annihilate a positron) 
and annihilate  left (right) handed current neutrino $\nu_{eL}$ 
($\nu_{eR}$). The weak interaction Hamiltonian (\ref{eq:5}) is given 
for $\zeta  \ll 1$ and $M^W_2 \gg M^W_1$ keeping only the lower order terms
in expansion of $tg \zeta$ and $M^W_1/M^W_2$ parameters. 

The left-right symmetric models  
allow us to explain the smallness of the neutrino mass
within the so called see-saw mechanism in the most
natural way. It is supposed that 
the neutrino mixing does take place according to
\begin{eqnarray}
\nu_{e L} &=& \sum_{k=light} ~U^L_{ek}~\chi_{kL} +
\sum_{k=heavy} ~U^L_{ek}~N_{kL}, \nonumber \\
\nu_{e R} &=& \sum_{k=light} ~U^R_{ek}~\chi_{kR} +
\sum_{k=heavy} ~U^R_{ek}~N_{kR},
\label{eq:6}   
\end{eqnarray}
where, $\chi_{k}$ ($N_{k}$)
are fields of light (heavy) Majorana neutrinos with masses
$m_k$ ($m_k << 1$ MeV) and $M_k$ ($M_k >> 100$ GeV), respectively, 
and $U^L_{ek}$, $U^R_{ek}$ are  unitary mixing matrices. 
In the case of the most general lepton mixing
originating from a Dirac-Majorana mass term in the Lagrangian
the flavor neutrino fields are superposition of three light 
and three heavy Majorana neutrinos with definite mass. 
The fields $\chi_k$ and $N_k$ satisfy the Majorana condition: 
$\chi_k \xi_k = C ~{\overline{\chi}}_k^T$, 
$N_k {\hat \xi}_k = C ~{\overline{N}}_k^T$,
where C denotes the charge conjugation and $\xi$, ${\hat \xi}$ 
are phase factors. 

The possible quark level neutrino mixing mechanisms
of $0\nu\beta\beta$-decay are displayed 
in Figs. \ref{fig.1} (a), (b) and (c).  If the 
$0\nu\beta\beta$-decay is triggered by exchange of a
light (heavy) left-handed Majorana neutrino [see Fig. \ref{fig.1} (a)]
the corresponding amplitude of the process is proportional
to the LN violating parameter $<m_\nu >$ 
($\eta_{_N}$):
\begin{eqnarray}
<m_\nu > ~ &=& ~ \sum^{light}_k~ (U^L_{ek})^2 ~ \xi_k ~ m_k, 
\nonumber \\
\eta_{_N} ~ &=& ~ \sum^{heavy}_k~ (U^L_{ek})^2 ~ 
{\hat \xi}_k ~ \frac{m_p}{M_k},
\label{eq:7}   
\end{eqnarray}
where $m_p$  is the proton  mass. The difference between 
$<m_\nu >$ and $\eta_{_N}$ comes from the fact that the neutrino
propagator in the first and second case show different dependence
on the mass of neutrinos \cite{fae98}. We note that  
even if the neutrino is a Majorana particle but massless 
(i.e. there is no mixing of neutrinos), the process in
Fig. \ref{fig.1} (a) can not happen since for a pure left-handed 
weak interaction theory, the emitted neutrino must be right-handed 
(positive helicity), while the absorbed neutrino must be left-handed
(negative helicity). 
With a finite mass the neutrino has not any more a good helicity
and the interference term between the leading helicity and the
small admixtures allows a $0\nu\beta\beta$-decay. 

The presence of the slight right-handed weak interaction 
in the GUT's allows the mechanisms 
drawn in Figs. \ref{fig.1} (b) and (c). In this cases there is no helicity 
matching problem. The emitted neutrino from the left-handed vertex 
is right-handed as well as the absorbed neutrino at the right-handed vertex. 
Assuming only light neutrinos we distinguish two cases. 
i) The chiralities of the quark hadronic 
currents are the same 
as those of the leptonic currents coupled through the 
W-boson propagator [Fig. \ref{fig.1} (b)]. Thus  
the $0\nu\beta\beta$-decay amplitude is proportional to 
\begin{equation}
<\lambda > ~ = ~ \left( \frac{M^W_1}{M^W_2}\right)^2~
\sum^{light}_k~ U^L_{ek}~U^R_{ek} ~ \xi_k.
\label{eq:8}
\end{equation}
Recall that the W-boson propagator can be approximated
by $1/M^2$ for $M = M^W_1, M^W_2$ 
[see Eq. (\ref{eq:4})].\\
ii) The chirality of the right-handed leptonic current is 
opposite to the 
coupled left-handed hadronic current [Fig. \ref{fig.1} (c)]. 
This is possible due to W-boson mixing. 
The corresponding effective LN violating
parameter is
\begin{equation}
<\eta > ~ = ~ tg \zeta
\sum^{light}_k~ U^L_{ek}~U^R_{ek} ~ \xi_k 
\label{eq:9}
\end{equation}
It is worthwhile to notice that the factor 
$\sum^{light}_k~ U^L_{ek}~U^R_{ek} ~ \xi_k$ in Eqs. 
(\ref{eq:8}) and (\ref{eq:9}) is expected to be
small and even in the case there are only light neutrinos 
vanish due to orthogonality condition \cite{bil87}.
It indicates that the values of $<\lambda >$ and
$<\eta >$ might be strongly suppressed assuming 
the see-saw neutrino mixing mechanism.

\subsection{Majoron mechanism}

The spontaneous breaking of the LN in the context
of the see-saw model imply the existence of a  physical 
Nambu-Goldstone boson  \cite{chi80}, 
called Majoron \cite{gel81}, which is a 
light or massless boson with a very tiny coupling to neutrinos 
\begin{equation}
{\cal L}_{\phi\nu\nu} = \sum_{i \le j} 
{\overline{\nu}}_{i} \gamma_5  \nu_j ~(i ~Im~ \phi ) ~P_{ij},
~~~~P_{i,j} = \sum_{\alpha , \beta = e,\mu ,\tau} 
U^R_{i\alpha} ~U^R_{j\beta} ~g_{\alpha \beta},
\label{eq:10}   
\end{equation}
Here, $\nu_i$ denotes both light
$\chi_i$ and heavy $N_i$ Majorana neutrinos. 

The Majoron $\phi$ might occur in the Majoron mode of the 
$0\nu\beta\beta$-decay ($0\nu\beta\beta\phi$-decay)
\begin{equation}
(A,Z) \rightarrow (A,Z+2) + 2e^- + \phi.
\label{eq:11}
\label{int.3}
\end{equation} 
and offers a new possibility
for looking for a signal of new physics in double beta decay 
experiments. The  $0\nu\beta\beta\phi$-decay mode yields
a continuous electron spectrum for the sum of electron energies 
like the $2\nu\beta\beta$-decay mode 
but differs from it by the position of the maximum
as different numbers of light particles are present in the final state. 
We remind that in the case of  $0\nu\beta\beta$-decay a peak is 
expected to be at the end of the electron-electron coincidence spectra.

The mechanism leading to a $0\nu\beta\beta\phi$-decay mode is drawn
in Fig. \ref{fig.1} (d). The experimental lower limits on the 
half-life of $0\nu\beta\beta\phi$-decay allow to deduce the upper
limit on the effective Majoron coupling constant $<g>$:
\begin{equation}
<g> = \sum_{i j}^{light} U^L_{ei} U^L_{ej} P_{ij}. 
\label{eq:12}
\end{equation}

\subsection{R-parity violating SUSY mechanism}

Besides the simplest and the best known  mechanism of LN 
violation based on the mixing of massive Majorana neutrinos advocated  
by different variants of the GUT's the R-parity 
violation proposed in the context of the MSSM 
is becoming the most popular scenario
for LN violation (see e.g. reviews \cite{BAR98,BFK98}).
We remind that the R-parity symmetry  assigns even R-parity 
to known particles of the SM and odd R-parity to their
superpartners and that the Lagrangian of the MSSM conserves R-parity.
The R-parity conservation  is not required by 
gauge invariance or supersymmetry and might be broken at the Planck scale.
The R-parity violation ($R_p \hspace{-1em}/\;\:$)  is introduced in
the effective Lagrangian (or superpotential) of the MSSM in terms
of a certain set of hidden sector fields. The trilinear part 
of the  $R_p \hspace{-1em}/\;\:$ superpotential takes the form 
\begin{equation}
W_{\Rslash} = \lambda_{ijk} L_i L_j E^c_k
           + \lambda_{ijk}' L_i Q_j D^c_k 
	   + \lambda_{ijk}''  U^c_i D^c_j D^c_k.
\label{eq:13}
\end{equation}
Here $L$ and  $Q$ stand for lepton and quark doublet left-handed
superfields while $E^c$,$U^c$, $D^c$ stand for lepton and $up$,
$down$ quark singlet superfields.
$\lambda_{ijk} $, $\lambda'_{ijk}$ and $\lambda''_{ijk}$
are coupling constant and 
indices $i, j, k$ denote generations. The $\lambda''$-terms are
causing baryon number violation and the remaining ones the 
LN violation. In fact a combination of 
$\lambda'$ and $\lambda''$ leads to proton decay.

\begin{figure}[t]
\centerline{\epsfig{file=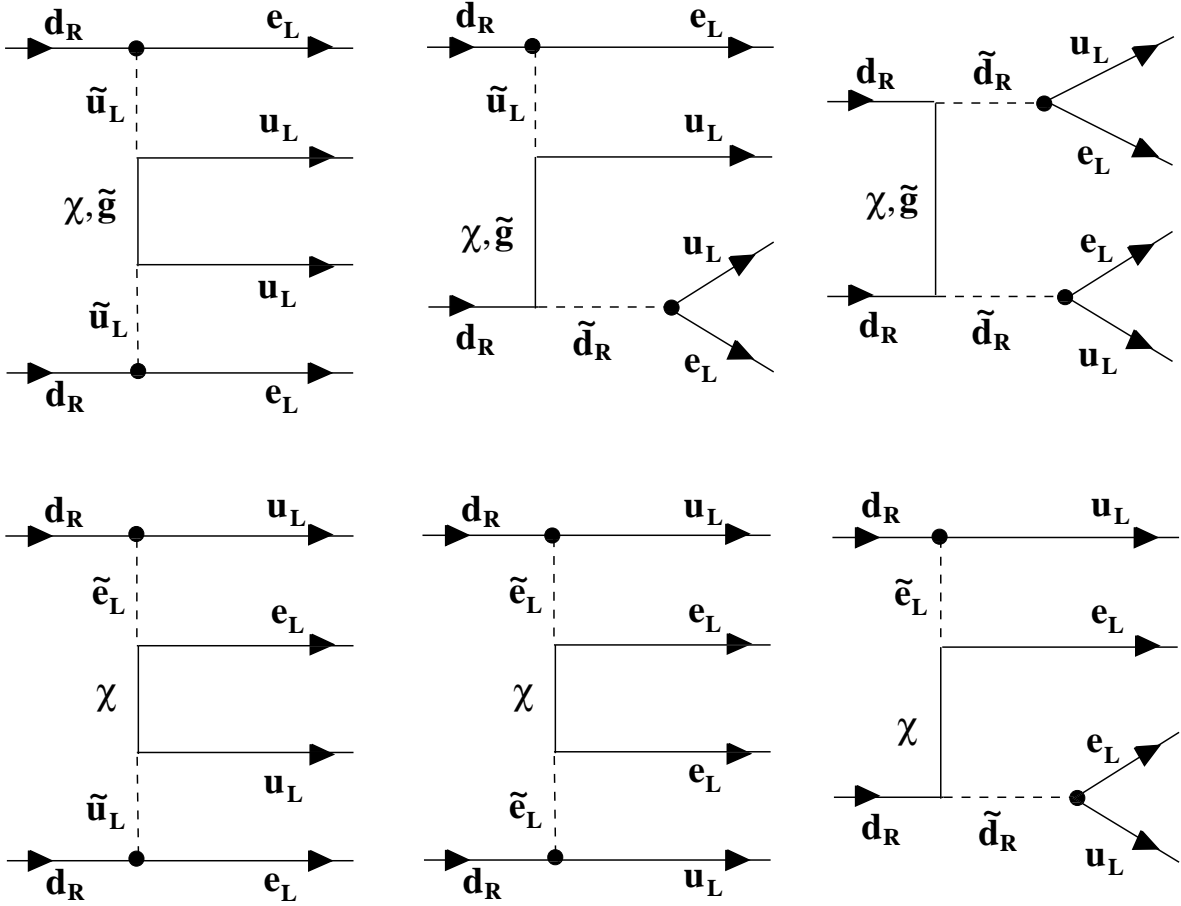,height=12.0cm}}
\caption{
Feynman graphs for the supersymmetric contributions to 
$0\nu\beta\beta$-decay. $u_L$, $d_R$ and $e_L$ have the same meaning 
as in Fig. \ref{fig.1}. 
${\tilde{u}}_L$, ${\tilde{d}}_R$ and ${\tilde{e}}_L$
are left-handed u-squark, right-handed d-squark and left-handed
selectron, respectively. $\chi$ and $\tilde{g}$ are neutralinos and
gauginos, respectively.
}
\label{fig.2}
\end{figure}

If the $R_p \hspace{-1em}/\;\:$ SUSY models are correct
the $0\nu\beta\beta$-decay is feasible. 
The nuclear ${0\nu\beta\beta}$-decay
is triggered by the ${0\nu\beta\beta}$-decay quark transition
$dd\rightarrow uu + e^- e^-$. The relevant Feynman diagrams 
associated with gluino ${\tilde g}$ and neutralino $\chi $ 
trilinear $R_p \hspace{-1em}/\;\:$ SUSY contributions 
to the $0\nu\beta\beta$-decay are drawn in Fig. \ref{fig.2}. 
The $R_p \hspace{-1em}/\;\:$  SUSY vertices are indicated with
bold points. We see that the $0\nu\beta\beta$-decay amplitude
is proportional to the $\lambda'_{111}$ squared.

There are two possibilities of the hadronization, i.e. coming
from the quark level to the nucleon level. 
One can place the four quark  into the
two initial neutrons and two final protons, what is just the
conventional two-nucleon mechanism of $0\nu\beta\beta$-decay
($nn \rightarrow pp + e^- e^-$). This mechanism 
is strongly suppressed by the nucleon-nucleon repulsion 
at short distances for the exchange of heavy SUSY particles
and heavy Majorana neutrinos. Another possibility is to 
incorporate quarks involved in pions in flight
between nucleons. This possibility 
have been first pointed out by Pontecorvo \cite{pon68}.
It is the so called pion exchange mechanism 
($\pi^- \rightarrow \pi^+  + e^- e^-$). The pion-exchange mode leads to 
a long-range nuclear interaction, which is significantly less 
sensitive to short--hand correlations effects. It was found that 
$R_p \hspace{-1em}/\;\:$ SUSY pion exchange contribution to the 
${0\nu\beta\beta}$-decay absolutely dominates over the 
conventional two nucleon mode realization \cite{FKSS97,awf99,verl99}. 

The enhancement of the pion-exchange mode has also an origin
in the bosonization of the $\pi^- -> \pi^+ + e^-e^-$ vertex and is
associated with the pseudoscalar $J_P J_P$ 
and tensor $J_T^{\mu\nu} J_{T\mu\nu}$ hadronic current structure  
of the effective R-parity violating $0\nu\beta\beta$-decay Lagrangian 
on the quark level \cite{FKSS97}
($J_P= \overline{u} \gamma_5 d $ and 
($J_T^{\mu\nu}= \overline{u} \sigma^{\mu\nu} (1+\gamma_5) d $). 
The corresponding hadronic matrix 
elements are given as follows:
\begin{eqnarray}
<\pi^+(q) | J_P J_P  | \pi^-(q) > &\approx&
\frac{5}{3} <\pi^+(q) | J_P |0> <0| J_P |\pi^-(q) > \nonumber \\ 
&=& - \frac{10}{9} f^2_\pi 
\frac{m^4_\pi}{(m_u +m_d)^2} = - m^4_\pi c_P, \nonumber \\
<\pi^+(q) | J_T^{\mu\nu} J_{T\mu\nu}  | \pi^-(q) > &\approx&
-4 <\pi^+(q) | J_P |0> <0| J_P |\pi^-(q) >.
\label{eq:14}
\end{eqnarray}
Here $m_\pi$ is the pion mass, $f_\pi = 0.668 ~m_\pi$. Taking
the conventional values of the current quark masses 
$m_u = 4.2$ MeV, $m_d = 7.4$ MeV one gets 
$c_P \approx 214.$ In the case of the exchange of a heavy Majorana
neutrino there is a vector and axial-vector hadronic current structure
$J_{AV}^\mu J_{AV\mu}$ of the effective Lagrangian 
($J^\mu_{AV} = {\bar u}^{\alpha} \gamma^\mu (1-\gamma_5) d_{\alpha}$). 
We have
\begin{eqnarray}
<\pi^+(q) | J_A^\mu J_{A\mu}  | \pi^-(q) > &\approx&
\frac{8}{3} <\pi^+ (q) | J_P |0> <0| J_P |\pi^- (q) > \nonumber \\ 
&=& - \frac{8}{3} f^2_\pi q^2 = - m^4_\pi c_A(q^2).
\label{eq:15}
\end{eqnarray}
Assuming the average momentum of the exchanged pion to be about
100 MeV we find $c_A \approx 0.61 $. We note that $c_P \gg c_A$.

In order to derive a limit on the R-parity violating first 
generation Yukawa coupling $\lambda_{111}'$ from the
observed absence of the $0\nu\beta\beta$-decay it is necessary 
to use viable 
phenomenological assumptions about some of the fundamental 
parameters of the $R_p \hspace{-1em}/\;\:$ MSSM.  
In Ref. \cite{FKSS97}) the ansatz of universal 
sparticle masses was assumed and that the lightest neutralino is
bino-like.  Within such phenomenological scenarios
it was found that the gluino and neutralino exchange mechanisms 
are of comparable importance \cite{fae98,FKSS97}. 
Another possibility is to implement
relations among the weak scale values of all parameters
entering the superpotential and the soft SUSY breaking Lagrangian and 
their values at the GUT scale. This scenario have been outlined 
in Refs. \cite{awf99,gauge1}. It was shown that 
there is no unique answer to the problem of the dominance of 
neutralino and gluino contribution to $0\nu\beta\beta$-decay.
The dominance of any of them is bound with a
different choice of the SUSY parameters $m_0$ and $m_{1/2}$. 
It is worthwhile to notice that in the extraction of $\lambda'_{111}$
the main uncertainty comes from the parameters of supersymmetry
and not from the nuclear physics side \cite{awf99,verl99}.

\section{$2\nu\beta\beta$-decay and nuclear structure}
\label{sec:level3}

Since there are
measurements available for the $2\nu\beta\beta$-decay with the 
geochemical method 
($^{82}Se$ \cite{lin88}, $^{96}Zr$ \cite{kaw93}, $^{128}Te$ and
$^{130}Te$ \cite{bern92}) and with the radiochemical method 
($^{238}U$ \cite{tur91} and for seven
nuclei even laboratory measurements 
( $^{48}Ca$ \cite{bal96}, $^{76}Ge$ \cite{gue97}, 
$^{82}Se$ \cite{arn98}, $^{96}Zr$ \cite{bara98,barp,fran}, 
$^{100}Mo$ \cite{sil97,das95}, $^{116}Cd$ \cite{dane95,arn96} and
$^{150}Nd$ \cite{sil97}), one could try to
calculate for a test of the theory the double beta-decay with two
neutrinos and compare them with the data. We note that a
positive evidence for a $2\nu\beta\beta$-decay transition to the 
$0^+_1$ excited state of final nucleus was observed for $^{100}Mo$
\cite{bexc}.

\begin{figure}[t]

\centerline{\epsfig{file=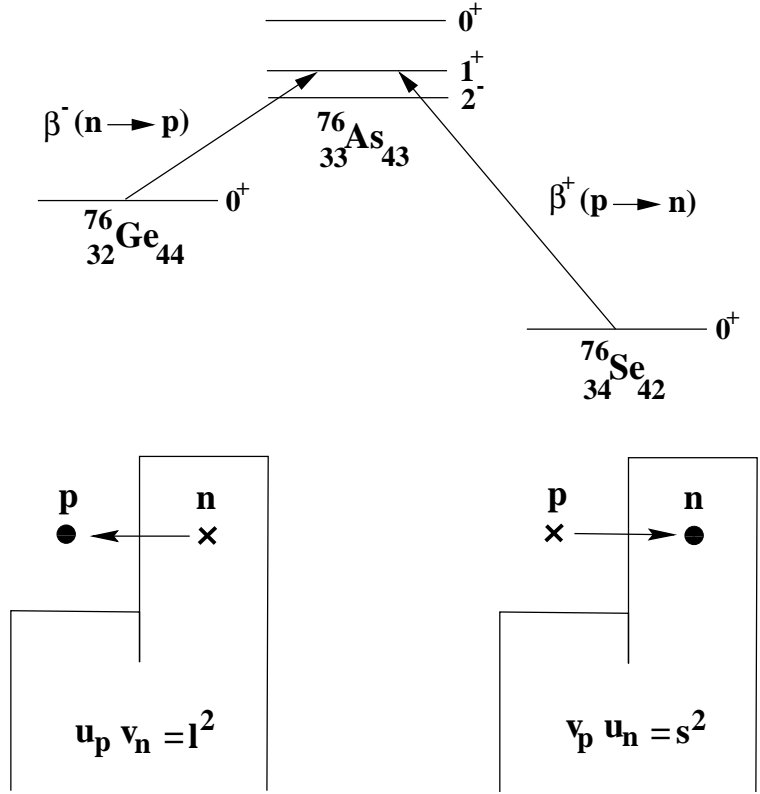,width=10.0cm}}
\caption{
The upper part shows the way how in the Random Phase
Approximation
(RPA) the $2\nu \beta \beta$ decay is calculated.
For the Fermi transitions the $\beta ^{-}
(n \to \ p)$ amplitude moves just a neutron into the same proton
level and the $\beta ^{+} (p \to \ n)$ amplitude moves a proton
into the same neutron level. For the Gamow-Teller transitions it
can also involve a spin flip, but the orbital part remains the
same. One immediately realizes that the occupation and
non-occupation amplitudes favor the $\beta ^{-}$ amplitude, but
disfavor the $\beta ^{+}$ amplitude. There one has a transition
from an unoccupied to an occupied single particle state, which is
two-fold small $(s^2)$ first by the fact that the occupation
amplitude for the proton $v_p$ and secondly that the unoccupation
amplitude for the neutron state $u_n$ are both small. Therefore
the $2\nu \beta \beta $ is drastically reduced.
}
\label{fig.3}
\end{figure}

The inverse half-life of the $2\nu\beta\beta$-decay is free of 
unknown parameters on the particle physics side and can
be expressed as 
a product of  a phase-space factor $G^{2\nu}$ and the Gamow-Teller
transition matrix element $M^{2\nu}_{GT}$ in second order:
\begin{equation}
[ T^{2\nu}_{1/2}(0^+_{g.s.} \rightarrow 0^+_{g.s.}) ]^{-1} = 
G^{2\nu} ~ | M^{2\nu}_{GT}|^2,
\label{two:16}
\end{equation}
where
\begin{eqnarray}
M^{2\nu}_{GT}&=&\sum_{{n}}\frac{
<0^{{+}}_{{f}}|A_{{k}}(0)|1^{{+}}_{{n}}>
<1^{{+}}_{{n}}|A_{{k}}(0)|0^{{+}}_{{i}}>}
{E_{{n}}-E_{{i}}+\Delta}.
\label{two:17}
\end{eqnarray}
$|0^{{+}}_{{i}}>$, $|0^{{+}}_{{f}}>$ 
and $|1^{{+}}_{{n}}>$ are respectively the 
wave functions of the initial, final and intermediate nuclei with
corresponding energies $E_{{i}}$, $E_{{f}}$ 
and $E_{{n}}$. $\Delta$ denotes
the average energy $\Delta = (E_{{i}}-E_{{f}})/2$. 
$A_{{k}}$ is the
Gamow-Teller transition operator $A_{{k}}=\sum_{{i}} 
\tau^{{+}}_{{i}}(\vec{\sigma}_{{i}})_{{k}}$, 
k=1,2,3.

The calculation of $M_{GT}$ remains to be challenging and attracts
the specialists of different nuclear models. The computational
complexity of $M_{GT}$ consists in the reliable description of the 
complete set of the intermediate nuclear states. Recently, 
it has been shown the summation over the intermediate nuclear states 
in the present $2\nu\beta\beta$-decay studies corresponds to a summation 
over a class of meson exchange diagrams within the S-matrix approach
\cite{sim99}. 

The nuclear shell model gives a satisfactory description
only of the low lying excited states of nuclei. In the heavier nuclei
there is a large number of basis states in the shell model which
does not allow to perform realistic calculation without severe truncations.
It is supposed to be the reason that the shell model predictions 
of $2\nu\beta\beta$-decay
rate for heavier nuclei, especially those for the Te region, show
deviations from  the experimental data \cite{vog99}.
We note that the feasibility of shell model calculations is growing 
with increasing computer facilities allowing to handle much larger
configuration spaces. 

Many different nuclear  structure aspects of the many-body Green function 
$M_{GT}$ have been discussed. It was suggested by Abad et al. \cite{abad}
that $M_{GT}$ could be dominated by the transitions through the
lowest intermediate $1^+$ state (so called 
Single--State--Dominance--Hypothesis (SSDH)). The SSDH could be 
realized in two ways:i) There is the true 
dominance of the first $1^+$ state, i.e. the contribution from 
higher lying $1^+$ states to $M_{GT}$ is negligible.
ii) There is a  cancelations among the contributions of higher lying $1^+$ 
states of the intermediate nucleus. 
The idea of SSDH have been outlined
in Ref. \cite{civ99} showing that some experimental and theoretical
evidence supports it for a few $2\nu\beta\beta$-decay systems. 

The difficulty of the calculation of $M_{GT}$ consists in the fact
that the $2\nu\beta\beta$-decay matrix elements are strongly 
suppressed. Its value is only small fraction of the double 
Gamow-Teller sum rule that scales roughly like the number of
pairs of unpaired neutrons \cite{eri88}, 
\begin{equation}
\sum_f |<f| \vec{A} \cdot \vec{A}|i>|^2 \approx 
(N-Z)(N-Z-1).
\label{eq:18}
\end{equation}
The proton-neutron QRPA  (pn-QRPA) has been found
successful in describing the suppression mechanism 
for the $2\nu\beta\beta$-decay \cite{vog86,civ87}. 
Fig. \ref{fig.3}  explains why the $2\nu\beta\beta$-decay amplitude is
so drastically reduced. Therefore the small effects which normally
do not play a major role can affect the $2\nu\beta\beta$-decay
transition probability. If one looks to the second leg of the
double beta-decay which is
calculated backwards as a $\beta^+ ~(p~ \rightarrow ~ n)$ 
decay from the final nucleus to the intermediate nucleus
one finds that the matrix elements involved in these diagrams 
are Pauli suppressed by a factor $(u_n v_p)^2
\ = \ (small)^4$. The neutron-particle proton-hole force in the isovector
channel, which is usually included is
repulsive while the particle-particle force usually neglected is
attractive. Therefore both excitations tend to cancel each other and
therefore the amplitude $\beta ^+$ is drastically reduced. 
This cancelation for the second leg could be even complete, i.e.
the backgoing amplitudes and thus groundstate
correlations cancel the leading forward going terms.

Although one can obtain agreement within pn-QRPA with the measured
$2\nu\beta\beta$ data multiplying the particle-particle G-matrix
elements of the nuclear Hamiltonian with a factor $g_{pp}$ 
in a range of $0.8 \leq g_{pp} \leq 1.2$
( $g_{pp}$ in principle should be equal to unity), two leaps of faith 
are usually quoted: i) The extreme sensitivity of $M_{GT}$ to 
the strength of particle-particle interaction which 
does not allow a reliable prediction of the 
$2\nu\beta\beta$-decay probability. We note that $M_{GT}$
as a function of $g_{pp}$ crosses zero.
ii) The collapse of the pn-QRPA solution within the physical
range of $g_{pp}$, what is supposed to be a phase transition. 
The collapse is caused by
generation of too many ground state correlations with increasing 
strength of the attractive proton-neutron interaction. 

The study of the QRPA approximation scheme for different model spaces
manifest that the problems i) and ii) are related \cite{simr88}
with each other. 
The undesirable behavior of the pn-QRPA has its origin in the quasiboson
approximation (QBA) violating the Pauli exclusion principle (PEP) 
and causing the QRPA excitation operators behave like bosons.
The renormalized QRPA, which considers the  PEP 
in an approximate way, shifts  the collapse of the QRPA outside 
the physical range of $g_{pp}$ and  shows a  less sensitive dependence 
of $M^{2\nu}_{GT}$ on $g_{pp}$  \cite{toi95,simn96}. It allows us to 
predict  more reliable values of the double beta decay matrix elements. 
The importance of the PEP for solving the
problem of the QRPA collapse has been shown clearly within the 
schematic models, which are trying to simulate the realistic cases
either by analytical solutions or by a minimal computational
effort \cite{sam97,fed99}. In Ref. \cite{fed99}, 
to our knowledge for the first time, the solution of the QRPA
equation with full consideration of the PEP was presented. 
It was found that restoring the PEP,  
the QRPA solutions are considerably stabilized and
a better agreement with the exact solution is obtained. 
A new extension of the standard pn-QRPA ``QRPA with PEP''
was proposed, which consider the PEP in more appropriate
way as the RQRPA and might work well also  in the
case of realistic calculations. 

We note that
the calculation of the  $2\nu\beta\beta$-decay nuclear transition 
continues to be subject of interest,  which stimulates the rapid 
development of the nuclear theory 
\cite{fed99,sam99,osv99,pas98,bob99,rad98,sco99}.

\section{Limits on LN violating parameters}
\label{sec:level4}

The limits deduced for LN violating parameters depend 
on the values of nuclear matrix element $ME^{0\nu}_i$, 
of the kinematical factor $G^{0\nu}_i$ and
of the current experimental limit for a given isotope
[see Eq. (\ref{eq:2})]. Thus there is useful
to introduce sensitivity parameters for a given isotope to 
the different LN violating parameters, which depend 
only on the characteristics of a given nuclear system.
There are the following:
\begin{eqnarray}
\label{eq:19}
\zeta_{<m_\nu >} (Y) & = &
10^{7}~ |M^{light}_{<m_\nu >}|~ 
\sqrt{{G_{01}}~ {year}},~~~~~~
\zeta_{\eta_{_N}} (Y)  =  
10^{6}~ |M^{heavy}_{<m_\nu >}|~ \sqrt{{G_{01}}~{year}},
\nonumber \\
\zeta_{<\lambda >} (Y) & = & 
10^{7}~ |M^{0\nu}_{GT}|~ 
\sqrt{C_{\lambda\lambda}~{year}},~~~~~~
\zeta_{<\eta >} (Y) = 
10^{5}~ |M^{0\nu}_{GT}|~ 
\sqrt{C_{\eta\eta}~{year}},\nonumber \\
\zeta_{<g>} (Y) & = & 
10^{8}~ |M^{light}_{<m_\nu >}|~ 
\sqrt{{G_{B}}~{year}},~~~~~~
\zeta_{\lambda_{111}'} (Y) = 
10^{5}~ |M^{\pi N}|~ 
\sqrt{{G_{01}}~{year}}.
\end{eqnarray}
The explicit form of 
$M^{light}_{<m_\nu >}$, $M^{heavy}_{<m_\nu >}$, 
$M^{0\nu}_{GT}$, $C_{\lambda\lambda}$, $C_{\eta\eta}$,
$C_{\lambda\lambda}$ and $M^{\pi N}$ can be found e.g. in 
Refs. \cite{pseu99,fae98,awf99}. 

Admittedly there is a rather large spread between the 
calculated values of nuclear matrix elements within different 
nuclear theories (for $^{76}Ge$ the calculated rates differ
by a factor of 3 \cite{fae98,suh98}). In principle there are no
exact criteria to decide which of them are correct. Nevertheless
one can argue the RQRPA method
offers more reliable results than the QRPA as the
ground state correlation are better under control due
to the consideration of PEP. 

The present limits on LN violating parameters $<m_\nu >$, $\eta_{_N}$,
$<\lambda >$, $<\eta >$ and $<g>$ are associated with
the two-nucleon mechanism for which the correct treatment of
the weak nucleon current $J^{\mu\dagger}$ is crucial. We have:
\begin{eqnarray}
J^{\mu \dagger}_L 
=  \overline{\Psi} \tau^+ \left[ g_V(q^2) \gamma^\mu 
- i g_M (q^2) \frac{\sigma^{\mu \nu}}{2 m_p} q_\nu
 - g_A(q^2) \gamma^\mu\gamma_5 + g_P(q^2) q^\mu \gamma_5 \right] \Psi,
\label{eq:20}   
\end{eqnarray}
where $q^\mu = (p-p')_\mu$ is the momentum transferred
from hadrons to leptons ($p$ and $p'$ are four momenta of neutron and 
proton, respectively) and 
$\sigma^{\mu\nu} = (i/2)[\gamma^{\mu}, \gamma^{\nu}]$.
$g_V(q^2)$, $g_M(q^2)$, $g_A(q^2)$ and $g_P(q^2)$ are 
the vector, weak-magnetism,
axial-vector and induced pseudoscalar formfactors, respectively,
which are real functions of a Lorentz scalar $q^2$. 

The matrix elements $M^{light}_{<m_\nu >}$ and $M^{heavy}_{<m_\nu >}$
have been calculated by neglecting the role of induced nucleon currents
(weak magnetism and induced pseudoscalar terms). Recently,
it has been shown that they contribute significantly to the
$0\nu\beta\beta$-decay amplitude \cite{pseu99}. They modify
the Gamow-Teller contribution and create a new tensor 
contribution. Their contribution is  as important as that of 
unchanged Fermi matrix element. It was found that indeed such
corrections cause a more or less uniform reduction of the 
$M^{light}_{<m_\nu >}$ by approximately $30\%$ throughout
the periodic table. In the case of heavy neutrino exchange 
($M^{heavy}_{<m_\nu >}$) the 
effect is much larger ( a factor of 3) \cite{pseu99}. 
The nucleon finite size has been 
taken into account through the phenomenological formfactors and
the PCAC hypothesis. We note that in 
calculating the matrix elements involving the 
exchange of heavy neutrinos, the treatment of the short-range 
repulsion and nucleon finite size is crucial \cite{lub92}. 
It is expected 
that the correct treatment of the induced pseudoscalar term (which is
equivalent to a modification of the 
axial-vector current due to PCAC) might influence significantly also the
amplitude of $0\nu\beta\beta$-decay in the case of the 
right-handed current mechanisms. It goes without saying that 
the validity of this argument
can be ultimatively assessed by numerical calculations. 

The numerical values of the sensitivity parameters $\zeta_X (Y)$
($X = <m_\nu >$, $\eta_{N}$, $<\lambda >$,
$<\eta >$,  ${<g>}$ and $\lambda_{111}'$) are listed
in Table \ref{table.1}. In their calculation we used 
values of $M^{light}_{<m_\nu >}$, $M^{heavy}_{<m_\nu >}$, 
and $M^{\pi N}$ calculated within the pn-RQRPA
\cite{pseu99,awf99}. As there are no available pn-RQRPA
results for $C_{\lambda\lambda}$, $C_{\eta\eta}$ we used those
of Ref. \cite{pan96}. The quantity $\zeta_X(Y)$ is an intrinsic 
characteristic of an isotope Y. The large numerical values of the 
sensitivity $\zeta_X(Y)$ correspond to those isotopes within the 
group of $\beta\beta$ decaying nuclei which are the most promising
candidates for searching for the LN violating parameter X. However,
we remind that there are also other microscopic and macroscopic properties
of the isotope, which are important for building a 
$0\nu\beta\beta$-decay detector. By glancing the Table \ref{table.1}
we see that the most sensitive isotope is $^{150}Nd$. It is mostly 
due to the large phase space integral and partial due to the 
larger nuclear matrix element \cite{pseu99,awf99}. We remark that
the nucleus $^{150}Nd$ is deformed and that in the calculation of the
corresponding nuclear matrix element the effects of nuclear deformation, 
which might be important, were not taken into account.

It is expected that 
the experimental constraints on the half-life of the $0\nu\beta\beta$-decay 
are expected to be more stringent in future. 
Knowing the values of  $\zeta_X(Y)$ there is a straightforward 
way to deduce limits on LN violating parameters X from the
experimental half-lives  $T^{0\nu -exp}_{1/2}$:
\begin{eqnarray}
\label{eq:21}
\frac{<m_\nu >}{m_e} &\leq& \frac{10^{-5}}{\zeta_{<m_\nu >}}
\sqrt{\frac{10^{24}~years}{T^{0\nu -exp}_{1/2}}}, ~~~~~~~
\eta_{_N} \leq \frac{10^{-6}}{\zeta_{\eta_{_N}}}
\sqrt{\frac{10^{24}~years}{T^{0\nu -exp}_{1/2}}}, \nonumber \\
{<\lambda >} &\leq& \frac{10^{-5}}{\zeta_{<\lambda >}}
\sqrt{\frac{10^{24}~years}{T^{0\nu -exp}_{1/2}}}, ~~~~~~~
<\eta > \leq \frac{10^{-7}}{\zeta_{<\eta >}}
\sqrt{\frac{10^{24}~years}{T^{0\nu -exp}_{1/2}}}, \nonumber \\
<g> &\leq& \frac{10^{-4}}{\zeta_{<g>}}
\sqrt{\frac{10^{24}~years}{T^{0\nu -exp}_{1/2}}}, \nonumber \\
(\lambda_{111}' )^2 &\leq&
\kappa^2~\left( \frac{m_{\tilde{q}}}{100~GeV}\right)^4~
\left(\frac{m_{\tilde{g}}}{100~GeV}\right)
\frac{10^{-7}}{\zeta_{\lambda_{111}'}}
\sqrt{\frac{10^{24}~years}{T^{0\nu -exp}_{1/2}}}.
\end{eqnarray}
$\kappa$ is equal to 1.8 
(gluino phenomenological scenario \cite{fae98}).
The normalization of $10^{24}$ years was chosen so that the 
$\zeta$'s are of order unity. 

The current experimental upper bounds on the $0\nu\beta\beta$-decay 
effective LN violating parameters of interest for different
isotopes are shown in Table \ref{table.1}. We see that the 
Heidelberg-Moscow experiment \cite{bau99} 
(In Table \ref{table.1}
we are giving the sensitivity of the experiment for $^{76}Ge$ being
$T^{exp-0\nu}_{1/2} \ge 1.6\times 10^{25}$ 
as we want to compare this value with those from other
experiments.)
offers the most restrictive limit for $<m_\nu >$, $\eta_{_N}$,
$<\lambda >$, $<\eta >$ and the  $^{128}Te$ experiment \cite{bern92}
for $<g>$. 
We note that if the experimental data from 
an other geochemical experiment on double beta decay of $^{128}Te$
would be considered
($T^{exp}_{1/2} = 1.5\times 10^{24}$, see Refs. cited in \cite{alle98}),
one would get less stringent limit on $<g>$ ($<g> \le 1.5\times 10^{-4}$),
which is comparable with the upper bound offered by
the $^{100}Mo$ experiment \cite{eji96} (see Table \ref{table.1}).

At present the largest attention is paid to the effective electron
Majorana neutrino mass parameter $<m_\nu >$ in light of the
positive signals in favor of oscillations of  solar,
atmospheric and terrestrial neutrinos. The masses and mixing angles can 
be determined from the available experimental data on neutrino
oscillations and from astrophysical arguments by using
some viable assumptions (hierarchical and non-hierarchical
neutrino spectra etc ).
 At present the three and four neutrino mixing 
patterns are the most favorable ones  \cite{mix99,oman99}. 
However, there is a discussion 
whether we really need the fourth sterile neutrino to fit 
the current experimental data. Knowing the elements of the
neutrino mixing matrix one can draw conclusions about the 
$0\nu\beta\beta$-decay, in particular on $<m_\nu >$, assuming
neutrinos to be Majorana particles. From the study of 
different neutrino mixing schemes it follows 
that the upper bound on effective Majorana neutrino mass $<m_\nu >$
could be ranging from $10^{-2}$ to 1 eV \cite{mix99,viss,barg}.
From the Table \ref{table.1} we see that the  
Heidelberg-Moscow $0\nu\beta\beta$-decay experiment \cite{bau99} implies
$<m_\nu >$ to be less 0.5 eV. This fact allows us to make
two important conclusions: i) The $0\nu\beta\beta$-decay
plays an important role in deciding among the alternative possibilities
of neutrino mixing. ii)  
The lepton number violation is in reach of near future 
$0\nu\beta\beta$-decay 
experiments, if the neutrino is a Majorana particle. 
An issue which only $0\nu\beta\beta$-decay can decide. 
We remind that the discovery of the $0\nu\beta\beta$-decay, what
would be a major achievement for particle physics
and cosmology, will implies only the upper bound on $<m_\nu >$ as
a plethora of other $0\nu\beta\beta$-decay mechanisms is in the
game. It is supposed that only further measurements of 
$0\nu\beta\beta$-decay transitions to the excited states of daughter
nucleus together with inclusion of nuclear theory and study
of different differential characteristics will allow us
to decide which mechanism is the dominant one. 

\begin{table}[t]
\caption{The present state of the Majorana neutrino mass (light and heavy),
right-handed current, Majoron and $R_p \hspace{-1em}/\;\:$ SUSY searches
in $0\nu\beta\beta$-decay experiments. 
$T^{exp-0\nu}_{1/2}$ and $T^{exp-0\nu \phi}_{1/2}$ 
are the best presently available lower limit on
the half-life of the $0\nu\beta\beta$-decay  and $0\nu\beta\beta\phi$-decay 
for a given isotope, respectively.
$\zeta_{X} (Y)$ denotes according to Eq. (19) the sensitivity of
a given nucleus $Y$ to the LN violating parameter X.
The  upper limits on ${<m_\nu >}$, $\eta_{N}$, $<\lambda >$,
$<\eta >$,  ${<g>}$ and $\lambda_{111}'$ are presented. 
gch.-geochemical data.}
\label{table.1}
\begin{tabular}{lrrrrrrrrr}\hline
Nucleus & $^{76}Ge$ & $^{82}Se$ & $^{96}Zr$ & $^{100}Mo$ &
 $^{116}Cd$ & $^{128}Te$ & $^{130}Te$ & $^{136}Xe$ & $^{150}Nd$ 
\\ \hline
$T^{exp-0\nu}_{1/2}$ & 
$ 1.6\times $ & 
$ 1.4\times $ &
$ 1.0\times $ & 
$ 2.8\times $ &  
$ 2.9\times $ &
$ 7.7\times $ &  
$ 5.6\times $ &
$ 4.4\times $ &  
$ 1.2\times $   \\
$[years]$  & 
$10^{25}$ & $10^{22}$ & $10^{21}$ & $10^{22}$ & $10^{22}$ & 
$10^{24}$ & $10^{22}$ & $10^{23}$ & $10^{21}$ \\
C.L. [\%] & 90 & 90 & 90 & 90 & 90 & gch. & 90  & 90 & 90 \\
Ref. & 
\cite{bau99} & \cite{ell92}  & \cite{fran}  & \cite{eji96}  &
\cite{dane95} & \cite{bern92} & \cite{alle98}  & \cite{bus98}  & 
\cite{sil97}   \\
 & & & & & & & & & \\
$T^{exp-0\nu\phi}_{1/2}$ &
$ 7.9\times $    & 
$ 2.4\times $    &
$ 3.9\times $    & 
$ 3.1\times $    &  
$ 1.2\times $    &
$ 7.7\times $    &  
$ 2.7\times $    &
$ 7.2\times $    &  
$ 2.8\times $   \\
$[years]$ &
$ 10^{21}$  & 
$ 10^{21}$  &
$ 10^{20}$  & 
$ 10^{21}$  &  
$ 10^{21}$  &
$ 10^{24}$  &  
$ 10^{21}$  &
$ 10^{21}$  &  
$ 10^{20}$   \\
C.L. [\%] & 90 & 90 & 90 & 90 & 90 & gch. & gch.  & 90 & 90 \\
Ref. & 
\cite{gue97} & \cite{arn98} & \cite{fran} & \cite{eji96} &
\cite{arn96} & \cite{bern92} & \cite{bern92} & \cite{bus98} & 
\cite{sil97}  \\
 & & & & & & & & & \\
$\zeta_{<m_\nu >} $ &  2.49 &  4.95 & 4.04 &  7.69 &  5.11 &
 1.02 &  4.24 & 1.60 &  17.3 \\
$\zeta_{\eta_{_N}} $  &
 2.90 &  5.64 & 3.98 &  7.10 &  5.36 &
 1.25 &  5.45 & 3.43 &  18.5 \\
 $\zeta_{\eta_{<\lambda >}} $  &
 3.35 & 6.92 & 10.3 & 1.81 & 1.60 & 0.66 & 8.62 & 6.06 & 27.6 \\
 $\zeta_{\eta_{<\eta >}} $  & 
 5.67 & 3.91 & 8.20 & 5.39 & 2.28 & 2.86 & 12.7 & 9.00 & 40.9 \\
$\zeta_{<g>} $  &
 2.41 &  6.59 & 5.93 & 10.5 &  6.60 &
 0.53 &  5.08 & 1.90 &  26.7 \\
 $\zeta_{\lambda_{111}'} $  &
 5.57 &  10.9 & 11.6 & 17.9 &  10.9 & 
 3.25 & 14.7 & 8.92 & 54.7 \\
 & & & & & & & & & \\
 $<m_\nu >~[eV]$  & 
0.51 & 8.7 & 40. & 4.0 & 5.9 & 1.8 & 5.1 & 4.8 & 8.5  \\
 $\eta_{_N}~[10^{-7}]$ & 
$0.86$ & $15.$ & $79.$ &
$8.4$ & $11.$ & $2.9$ & 
$7.7$ & $4.4$ & $16.$ \\   
$<\lambda >~[10^{-6}]$ &
 0.75 & 12. & 31. & 33. & 36. & 5.5 & 4.9 & 2.5 & 10.4 \\
$<\eta >~[10^{-8}]$ &
 0.44 & 22. & 38. & 11. & 26. & 1.3 & 3.3 & 1.7 & 7.1 \\
 $<g> ~[10^{-4}]$ & 
$4.7$ & $3.1$ & $8.5$ &
$1.7$ & $4.4$ & $0.69$ & 
$3.8$ & $6.2$ &  $2.2$ \\   
$\lambda_{111}'~(100~GeV)~[10^{-4}]$  &
1.2 & 5.0 & 9.4 & 3.3 & 4.2 & 
1.9 & 3.0 & 2.3 & 4.1 \\
$\lambda_{111}'~(1~TeV)~[10^{-2}]$  &
3.8 & 16. & 30. & 10. & 13. & 6.0 & 9.6 & 7.4 & 13. \\\hline
\end{tabular}
\end{table}

The present neutrino experiments does not provide us
with useful information about the mixing of heavy Majorana
neutrinos ($M_k >> 100$ GeV). Therefore, it is difficult to extract 
the mass of  heavy neutrinos from the current limit
$\eta_{_N} \le 8.6\times 10^{-8}$ (see Table \ref{table.1}). 
If one assumes the
corresponding $U_{ei}$ to be of the order of unity for the lightest
heavy neutrino mass $M_1$ one gets: $M_1 \ge 1.1\times 10^4$ TeV. 
However, this element of the neutrino mixing matrix is expected
to be rather small due to large differences in masses of light and heavy
neutrinos within the see-saw mechanism. Therefore the real limit on
$M_k$ is supposed to be much weaker. It is worthwhile to notice that
the limit on $\eta_{_N}$ is extremely sensitive to the nucleon finite size
and the short--range correlation effects \cite{lub92}. The 
heavy neutrino exchange nuclear
matrix elements evaluated without inclusion of the induced 
nucleon currents  \cite{opan96} are considerably overestimated \cite{pseu99}
and should not be used in deducing the limit on $\eta_{_N}$.

The present particle physics phenomenology does
not allow us to deduce  the magnitude of 
$\sum^{light}_k~ U^L_{ek}~U^R_{ek} ~ \xi_k$ 
entering the expressions for the effective right-handed current parameters
$<\lambda >$ and $<\eta >$. 
If we assume its value is about unity, then we get 
from the current limits on $ |<\lambda >| \le 7.5\times 10^{-7}$ and
$|<\eta >| \le 4.4\times 10^{-9}$ (see Table \ref{table.1}) the 
corresponding bounds on the mass $M_2^W$ of the 
heavy vector boson $W_2^\pm$ and
the mixing angle $\zeta$  as follows:
$M^W_2 \ge 93.$ TeV and
$|tg \zeta | \le 4.4\times 10^{-9}$. Mohapatra has shown that 
in the left-right symmetric model with spontaneous R-parity violation
there is an upper limit on $M_{W_R}$ (in the limit 
$\tan \zeta \rightarrow 0$ $M_{W_R} = M^W_2$) of at most 10 TeV
\cite{moh20}. By using this value for upper bound on
$M^W_2$ (for the lower limit on
$M^W_2$ we consider the value 100 GeV) we find
\begin{equation} 
| \sum^{light}_k~ U^L_{ek}~U^R_{ek} ~ \xi_k|  \le 1.1\times 10 ^{-6}~-~
1.1\times 10^{-2}, 
\end{equation} 
\begin{equation} 
|tg \zeta | \le 3.8\times 10^{-7} ~-~ 3.8\times 10^{-3}.
\end{equation} 
We remark that these limits could be modified after the pseudoscalar
term of the nuclear current is properly taken into account. 

The $0\nu\beta\beta$-decay offer the most stringent limit 
on the R-parity violating first 
generation Yukawa coupling $\lambda_{111}'$ \cite{BFK98}. 
Its value depends on the
masses of SUSY particles (see Eq. (\ref{eq:21})). If the masses of squark
$m_{\tilde{q}}$ and gluino $m_{\tilde{g}}$ would be at their present
experimental lower bounds of 100 GeV we deduce from the 
observed absence of the $0\nu\beta\beta$-decay 
$\lambda'_{111} \le 1.2 \times 10^{-4}$ (phenomenological scenario). 
A conservative upper bound is
obtained using the SUSY "naturalness" upper bound
$m_{\tilde{q}}$, $m_{\tilde{g}} \approx 1 $ TeV. It gives
$\lambda'_{111} \le 3.8 \times 10^{-2}$ (see Table \ref{table.1}). 
We mention that the limits 
on $\lambda'_{111}$ depend only weakly on the details of the 
nuclear structure as $\lambda'_{111}$ 
is proportional to the inverse square root of the nuclear matrix element.
In addition,  the corresponding nuclear matrix 
elements are changing only slightly within the physical range of 
parameters of the nuclear Hamiltonian \cite{awf99}. However, 
$\lambda'_{111}$ depends quadratic on the masses of SUSY particles. 
In the GUT's constrained SUSY scenario there is a rather large
SUSY parameter space.  By using different sample of relevant SUSY
parameters one ends up with significantly different limits on
$\lambda'_{111}$ \cite{verl99}. Finally, we stress that the above
limits are very strong and lie beyond the reach of near
future accelerator experiments (HERA, TEVATRON) 
\cite{FKSS97}. However, we note that the collider experiments  
are potentially sensitive to other couplings 
$\lambda'_{ijk}$, $\lambda''_{ijk}$ etc.

There are new experimental proposals for measurements of the
$0\nu\beta\beta$-decay for different isotopes. A new 
NEMO 3 experiment is under construction, which 
is expected to reach a lower limit on the $0\nu\beta\beta$-decay
half-life  of the order of $10^{25}$ years for 
$^{82}Se$, $^{96}Zr$, $^{100}Mo$, $^{116}Cd$, $^{130}Te$
and $^{150}Nd$ nuclei in a period of about six years \cite{barp}. 
The KAMLAND \cite{kl99}, CUORE \cite{crem} and 
GENIUS \cite{kl99} experiments are under consideration
at moment. The KAMLAND experiment is supposed to use as 
double beta decay emitter $^{136}Xe$ isotope in a liquid scintilator
and measure a half-life limit of about $5\times 10^{25}$ years.
In the CUORE experiment the cryogenic detector set-up will be made with 
crystals of $TeO_2$. The expected half-life limit for the
$0\nu\beta\beta$-decay of $^{130}Te$ could reach the value 
$1\times 10^{26}$ years. The largest half-life  limit of
$5.8\times 10^{27}$ years may be achieved in the future 
GENIUS experiment by using  one ton of enriched $^{76}Ge$ 
and one year for the measurement \cite{kl99}. If the above
experiments could be realized one would get considerably 
stronger limits on $0\nu\beta\beta$-decay lepton number
violating parameters. They are listed in the Table \ref{table.2}.
By glancing the Table  \ref{table.2} we see that NEMO 3 ($^{150}Nd$)
and CUORE ($^{130}Te$) experiments
have a chance to reach the value of 0.1 eV ($^{150}Nd$) for the 
effective neutrino mass and the GENIUS experiment could surpass
this border to lower value of 0.03 eV.

\begin{table}[t]
\caption{The expected limits on LN violating parameters
of interest from the future $0\nu\beta\beta$-decay experiments. 
The same notations as in Table I is used.}
\label{table.2}
\begin{tabular}{lccccccccc}\hline
exper. & nucl. & $T^{exp-0\nu}_{1/2}$ & $T^{exp-0\nu\phi}_{1/2}$ &
 $<m_\nu >$  & $\eta_{_N}$ & $<\lambda >$ & $<\eta >$ &
 $<g>$ & $\lambda_{111}'$  \\ 
 & & $10^{25}$ & $10^{23}$ & $[eV]$  & $[10^{-8}]$ & 
$[10^{-7}]$ & $[10^{-9}]$ & $[10^{-5}]$ & $[10^{-5}]$  \\ \hline
current & $^{76}Ge$  & 1.6 &     & 0.51  & 8.6 & 7.5 & 4.4 &     & 12. \\ 
        & $^{128}Te$ &     & 77. &       &     &     &     & 6.9 &     \\ 
NEMO-3  & $^{82}Se$  & 1.  & 1.  & 0.33  & 5.6 & 4.6 & 8.1 & 4.8 & 9.7 \\
        & $^{96}Zr$  & 1.  & 1.  & 0.40  & 7.9 & 3.1 & 3.8 & 5.3 & 9.4 \\
        & $^{100}Mo$ & 1.  & 1.  & 0.21  & 4.4 & 17. & 5.9 & 3.0 & 7.6 \\
        & $^{116}Cd$ & 1.  & 1.  & 0.32  & 5.9 & 20. & 14. & 4.8 & 9.7 \\
        & $^{130}Te$ & 1.  & 1.  & 0.38  & 5.8 & 3.7 & 2.5 & 6.2 & 8.3 \\
        & $^{150}Nd$ & 1.  & 1.  & 0.093 & 1.7 & 1.1 & 0.77 & 1.2 & 4.3 \\
KAMLAND & $^{136}Xe$ & 5.  &     & 0.45  & 4.1 & 2.3 & 1.6  &     & 7.2 \\
CUORE   & $^{130}Te$ & 10. &     & 0.12  & 1.8 & 1.2 & 0.79 &     & 4.7 \\
GENIUS  & $^{76}Ge$  & 580.&     & 0.027 & 0.45 & 0.39 & 0.23 &  & 2.8 \\
\hline
\end{tabular}
\end{table}

\section{Summary}
\label{sec:level5}

In this contribution we have discussed the problem of lepton
number violation in the context of rare nuclear processes,
in particular of the $0\nu\beta\beta$-decay, which has a broad 
potential for providing important information 
on modern particle physics.   We have shown
that the $0\nu\beta\beta$-decay
has strong impact on physics beyond the Standard model in the way it
constrains the parameters of other theories. 
The mechanism of LN violation within the $0\nu\beta\beta$-decay
 has been discussed in the context
of the problem of neutrino mixing  and the R-parity violating 
SUSY extensions of the Standard model.  The relevant LN
violating parameters have been  
the effective Majorana neutrino mass $<m_\nu >$, 
effective heavy neutrino mass parameter $\eta_{_N}$,
effective right-handed weak
interaction parameters $<\lambda >$ and $<\eta >$, 
effective Majoron coupling constant $<g>$ and
the first generation trilinear R-parity violating SUSY coupling
$\lambda_{111}$. The restrictions on the 
lepton number violating parameters have be deduced from the  
current experimental
constraints on $0\nu\beta\beta$-decay half-life time
for several isotopes of interest.

The present limit on the effective Majorana neutrino mass
$<m_\nu > \le 0.5 $ eV deduced from the $0\nu\beta\beta$-decay
experiment have been discussed in connection
with the different neutrino mixing scenarios advocated
by current data of neutrino experiments. We conclude that
the $0\nu\beta\beta$-decay experiment plays an important 
role in the 
determination of the character of the neutrino mass spectrum. 
Some analysis in  respect to the heavy Majorana neutrino 
have been also presented.
By using the upper and the lower limit on the mass of the heavy 
vector boson constrained by the 
left-right symmetric model with spontaneous R-parity violation
we have determined the allowed range for the
mixing angle of vector bosons. It has been found that
the current upper bound for the R-parity violating SUSY interaction
constant $\lambda'_{111}$  is
$\le 1.2 \times 10^{-4}$ ($\le 3.8 \times 10^{-2}$) assuming
the masses of SUSY particle to be on the scale of 100 GeV (1 TeV).
A discussion of the dominance of the pion-exchange 
R-parity violating mode for the $0\nu\beta\beta$-decay process
was also presented.

The interpretation of the LN violating
parameters involves some nuclear physics.  It is necessary to
explore the nuclear part of the $0\nu\beta\beta$-decay probability. 
The predictive power of different nuclear wave functions
can be tested in the $2\nu\beta\beta$-decay.
One needs a good description of the experimental data for the 
$2\nu\beta\beta$ probability. 
We have discussed the recent progress in the field of
the calculation of double beta decay matrix elements 
associated with the inclusion of the Pauli exclusion
principle in the QRPA. The reliability of the calculated
$0\nu\beta\beta$-decay matrix elements 
was addressed.

We found it useful to introduce sensitivity parameters 
$\zeta_X(Y)$
for a given isotope Y associated with different LN 
violating signals, which are free of influences from
particle physics
and relate simply the experimental half-lives
with LN violating parameters. The largest value of 
$\zeta_X(Y)$ determines that isotope, which is the most
sensitive to a given lepton number violating parameter 
X. It is an important information for planning new 
$0\nu\beta\beta$-decay experiments.

The $0\nu\beta\beta$-decay offers with both 
theoretical and experimental investigations a view to physics
 beyond the SM. New 
$0\nu\beta\beta$-decay experiments are in preparation or
under consideration (NEMO 3, KAMLAND, CUORE, GENIUS). 
They could verify the validity of different mixing scheme 
of neutrinos. The expected limits on the LN violating
parameters which could be reached in these 
experiments are presented in Table \ref{table.2}.
However, there is a possibility that the $0\nu\beta\beta$-decay 
could be detected in the forthcoming experiments.
This would establish that
the neutrino is a massive Majorana particle.
The recent development in neutrino physics has triggered 
the hope that we could be close to this achievement.

\end{document}